\documentclass[preprint,showpacs,preprintnumbers,amsmath,amssymb]{revtex4}
\usepackage{epsfig}
\usepackage{graphicx}% Include figure files
\usepackage{dcolumn}% Align table columns on decimal point
\usepackage{bm}% bold math
\usepackage{threeparttable}

\def\beq{\begin{equation}}
\def\eeq{\end{equation}}
\def\eeqn{\end{equation}}
\newcommand\iden{\leavevmode\hbox{\small1\normalsize\kern-.33em1}}

\newcommand{\bea} {\begin{eqnarray}}
\newcommand{\eea} {\end{eqnarray}}

\newcommand{\nn}{\nonumber}

\def\half{\tfrac{1}{2}}

\def\lamtil{\lambda_{345}}

\def\sb  {s_{\beta}}
\def\cb  {c_{\beta}}

\newcommand{\SLASH}[2]{\makebox[#2ex][l]{$#1$}/}
\newcommand{\eslash}{\SLASH{E}{.2}}

\let\jnfont=\rm
\def\NPB#1,{{\jnfont Nucl.\ Phys.\ B }{\bf #1},}
\def\PLB#1,{{\jnfont Phys.\ Lett.\ B }{\bf #1},}
\def\EPJC#1,{{\jnfont Eur.\ Phys.\ Jour.\ C }{\bf #1},}
\def\PRD#1,{{\jnfont Phys.\ Rev.\ D }{\bf #1},}
\def\PRL#1,{{\jnfont Phys.\ Rev.\ Lett.\ }{\bf #1},}
\def\MPLA#1,{{\jnfont Mod.\ Phys.\ Lett.\ A }{\bf #1},}
\def\JPG#1,{{\jnfont J.\ Phys.\ G }{\bf #1},}
\def\CTP#1,{{\jnfont Commun.\ Theor.\ Phys.\ }{\bf #1},}
\def\JHEP#1,{{\jnfont JHEP \ }{\bf #1},}
\def\NPPS#1,{{\jnfont Nucl.\ Phys.\ Proc.\ Suppl.\ }{\bf #1},}
\def\CPC#1,{{\jnfont Comput.\ Phys.\ Commun.\ }{\bf #1},}
\def\CPL#1,{{\jnfont Chin.\ Phys.\ Lett. }{\bf #1},}
\def\APPB#1,{{\jnfont Acta\ Phys.\ Polon.\ B }{\bf #1},}

\def\lsim{\raise0.3ex\hbox{$<$\kern-0.75em\raise-1.1ex\hbox{$\sim$}}}
\def\gsim{\raise0.3ex\hbox{$>$\kern-0.75em\raise-1.1ex\hbox{$\sim$}}}
\def\PR#1,{{\jnfont Phys.\ Rept. }{\bf #1},}
\def\CHC#1,{{\jnfont Chin.\ Phys.\ C }{\bf #1},}
\def\ASAS#1,{{\jnfont Astron.\ Astrophys. }{\bf #1},}
\begin{document}

\title{\ \\[10mm] Implication of the 750 GeV diphoton resonance on two-Higgs-doublet model and its extensions with Higgs field}
\author{Xiao-Fang Han$^{1}$, Lei Wang$^{2,1}$}
 \affiliation{$^1$ Department of Physics, Yantai University, Yantai
264005, P. R. China\\
$^2$ IFIC, Universitat de Val$\grave{e}$ncia-CSIC, Apt. Correus
22085, E-46071 Val$\grave{e}$ncia, Spain}

%---------------------------------------------------------------------------

\begin{abstract}
We examine the implication of the 750 GeV diphoton resonance on the
two-Higgs-doublet model imposing various theoretical and
experimental constraints. The production rate of two-Higgs-doublet
model is smaller than the cross section observed at the LHC by two
order magnitude. In order to accommodate the 750 GeV diphoton resonance, we
extend the two-Higgs-doublet model by introducing additional Higgs
fields, and focus on two different extensions, an inert complex
Higgs triplet and a real scalar septuplet. With the 125 GeV Higgs being agreement
with the observed data, the production rate for
the 750 GeV diphoton resonance can be enhanced to 0.6 fb for the
former and 4.5 fb for the latter. The results of the latter are well
consistent with the 750 GeV diphoton excess at the LHC.

\end{abstract}

\maketitle

\section{Introduction}
Very recently, the ATLAS and CMS collaborations have reported an
excess of events in the diphoton channel with an invariant mass of
about 750 GeV \cite{750}. The local significance of this signal is
at the $3\sigma$ level for ATLAS and slightly less for CMS. The
approximate production cross section times branching ratio is
4.47$\pm$1.86 fb for CMS and 10.6$\pm$2.9 fb for ATLAS by the
combination of 8 and 13 TeV data \cite{1512.04939}. However, there
are no excesses for the dijet \cite{dijet},
 $t\bar{t}$ \cite{ditt}, diboson or dilepton channels, which gives a challenge to the
the possible new physics model accommodating the 750 GeV diphoton
resonance. Some plausible explanations of this excess have already
appeared \cite{1512.04939,750work1,750work2,1512.04921}. The
two-Higgs-doublet model (2HDM) can not produce the enough large
cross section to accommodate the 750 GeV diphoton resonance.  Ref.
\cite{1512.04921} introduces some additional vectors-like quarks and
leptons to 2HDM in order to enhance the production rate of 750 GeV
diphoton resonance.

In this paper, we first examine the implication of the 750 GeV
diphoton resonance on the two-Higgs-doublet model imposing
various theoretical and experimental constraints. We give the
allowed mass ranges of the pseudoscalar and charged Higgs for
$m_H$ = 750 GeV, and find that the production rate of 2HDM is
smaller than the cross section observed at the LHC by two order
magnitude. Finally, in order to explain the 750 GeV diphoton excess,
we extend the two-Higgs-doublet model by introducing additional
Higgs fields, and focus on two different extensions, an inert
complex Higgs triplet and a real scalar septuplet. We find that the
production rate for the 750 GeV diphoton resonance can reach 0.6 fb
for the former and 4.5 fb for the latter with the 125 GeV Higgs
being consistent with the observed data.

Our work is organized as follows. In Sec. II we recapitulate the
two-Higgs-doublet model. In Sec. III we introduce the numerical
calculations, and examine the implications of the 750 GeV diphoton resonance
on the 2HDM after imposing the theoretical and experimental constraints.
 In Sec. IV, we respectively add an inert complex Higgs triplet and
a real scalar septuplet to 2HDM, and discuss the production rate for the 750 GeV diphoton resonance.
Finally, we give our conclusion in Sec. V.

\section{two-Higgs-doublet model}
The general Higgs potential is written as \cite{2h-poten}
\begin{eqnarray} \label{V2HDM} \mathrm{V} &=& m_{11}^2
(\Phi_1^{\dagger} \Phi_1) + m_{22}^2 (\Phi_2^{\dagger}
\Phi_2) - \left[m_{12}^2 (\Phi_1^{\dagger} \Phi_2 + \rm h.c.)\right]\nonumber \\
&&+ \frac{k_1}{2}  (\Phi_1^{\dagger} \Phi_1)^2 +
\frac{k_2}{2} (\Phi_2^{\dagger} \Phi_2)^2 + k_3
(\Phi_1^{\dagger} \Phi_1)(\Phi_2^{\dagger} \Phi_2) + k_4
(\Phi_1^{\dagger}
\Phi_2)(\Phi_2^{\dagger} \Phi_1) \nonumber \\
&&+ \left[\frac{k_5}{2} (\Phi_1^{\dagger} \Phi_2)^2 + \rm
h.c.\right]+ \left[k_6 (\Phi_1^{\dagger} \Phi_1)
(\Phi_1^{\dagger} \Phi_2) + \rm h.c.\right] \nonumber \\
&& + \left[k_7 (\Phi_2^{\dagger} \Phi_2) (\Phi_1^{\dagger}
\Phi_2) + \rm h.c.\right].
\end{eqnarray}
Here we focus on the CP-conserving case where all
$k_i$ and $m_{12}^2$ are real, and take $k_6=k_7=0$.
 This can be realized by introducing a discrete $Z_2$
symmetry, and $m_{12}^2$ is a soft-breaking term. The two complex
scalar doublets have the hypercharge $Y = 1$,
\begin{equation}
\Phi_1=\left(\begin{array}{c} \phi_1^+ \\
\frac{1}{\sqrt{2}}\,(v_1+\phi_1^0+ia_1)
\end{array}\right)\,, \ \ \
\Phi_2=\left(\begin{array}{c} \phi_2^+ \\
\frac{1}{\sqrt{2}}\,(v_2+\phi_2^0+ia_2)
\end{array}\right).
\end{equation}
Where the electroweak vacuum expectation values (VEVs) $v^2 = v^2_1
+ v^2_2 = (246~\rm GeV)^2$, and the ratio of the two VEVs is defined
as usual to be $\tan\beta=v_2 /v_1$. After spontaneous electroweak
symmetry breaking, there are five physical Higgses: two neutral
CP-even $h$ and $H$, one neutral pseudoscalar $A$, and two charged
scalar $H^{\pm}$.

The general Yukawa interaction can be given as
\begin{eqnarray}
{\cal L}_Y &=& \frac{-1}{\sqrt{2}}
 \bar{f}  \left[ z^f\sin(\beta-\alpha)+\rho^f\cos(\beta-\alpha) \right] h^0 f \nonumber\\
&&+\frac{-1}{\sqrt{2}} \bar{f}
 \left[ z^f\cos(\beta-\alpha)-\rho^f\sin(\beta-\alpha) \right] H^0  P_R f
 +\frac{i}{\sqrt{2}} \, {\rm sign}(Q_f)\bar{f} \rho^f A^0 P_R f\nonumber \\
 && -\bar{u} \left[ V \rho^d P_R - \rho^{u\dagger} V P_L \right] d H^+
  -\bar{\nu} \left[ \rho^\ell P_R \right] \ell H^+ + {\rm H.c.}.
\end{eqnarray}
Where $f=u~,d,~\ell$, and $z^f = \sqrt{2}m_f/v$, while $\rho$ matrices are free and have both
 diagonal and off-diagonal elements. For the aligned 2HDM \cite{a2hm}, $\rho=\sqrt{2}m_f\kappa_f/v$, which
leads that the couplings of neutral Higgs bosons
normalized to the SM Higgs boson are give by \bea &&
y^h_{V}=\sin(\beta-\alpha),~~~y^h_f=\sin(\beta-\alpha)+\cos(\beta-\alpha)\kappa_f,\nonumber\\
&&y^H_{V}=\cos(\beta-\alpha),~~~y^H_f=\cos(\beta-\alpha)-\sin(\beta-\alpha)\kappa_f,\nonumber\\
&&y^A_{V}=0,~~~~~~~y^A_u=-i\gamma^5
\kappa_{u},~~~~~~~~y^A_{d,\ell}=i\gamma^5 \kappa_{d,\ell}.\eea Where
$V$ denotes $Z$ and $W$. We fix $\kappa_u=1/\tan\beta$, which denotes that
a special basis is taken where there is no the up-type quark coupling to
$\Phi_1$ \cite{a2hm2,a2hfree}.

\section{numerical calculations and results of 2HDM}
\subsection{numerical calculations}
We use $\textsf{2HDMC}$ \cite{2hc-1} to implement the
theoretical constraints from the vacuum stability, unitarity and
coupling-constant perturbativity \cite{2hpert}, and calculate the oblique
parameters ($S$, $T$, $U$) and $\delta\rho$. $\textsf{HiggsBounds-4.1.4}$ \cite{hb-1,hb-2} is
employed to implement the exclusion constraints from the neutral and
charged Higgses searches at LEP, Tevatron and LHC at 95\% confidence
level. The in-house code is used to calculate the $B\to X_s\gamma$, $\Delta m_{B_s}$,
and $R_b$. The experimental values of electroweak precision data, $B\to X_s\gamma$,
$\Delta m_{B_s}$ are taken from \cite{pdg2014} and $R_b$ from
\cite{rb-exp}.

Since we focus on the implications of 750 GeV diphoton resonance on
the 2HDM, we fix $m_h=125$ GeV, $m_H=$
750 GeV, $|\sin(\beta-\alpha)|=1$, $\kappa_d=\kappa_\ell$=0. The last three
choices can naturally accommodate the non-observation of excesses
for diboson, dijet and dilepton. We scan randomly the parameters in
the following ranges:
\begin{eqnarray}
&&375 {\rm\  GeV} \leq m_{H^\pm}  \leq 2000  {\rm\  GeV},\nn\\
&& 0.5 \leq \tan \beta \leq 5,\nn\\
&&-(2000~{\rm GeV})^2 \leq m^2_{12} \leq (2000~{\rm GeV})^2.
\end{eqnarray}
Since the heavy CP-even Higgs coupling to the top quark is
proportional to $1/\tan\beta$ for $\cos(\beta-\alpha)=$0, we take the small $\tan\beta$ to
avoid the sizable suppression of this coupling. We take $m_A\simeq
m_H$ which leads that the 750 GeV diphoton resonance is from both
 $H$ and $A$, and the more large cross section may be
obtained. We define \beq R_{\gamma\gamma}\equiv\sigma(gg\to H)\times
Br(H\to \gamma\gamma)+\sigma(gg\to A)\times Br(A\to \gamma\gamma).
\eeq In this paper we will introduce additional Higgs fields to
2HDM, and some multi-charged scalars give very important
contributions to the CP-even Higgs decay into $\gamma\gamma$.
Therefore, we give the general formulas for the CP-even Higgs decay
into $\gamma\gamma$ \cite{hrr1loop},
    \beq \label{hrrgs}
    \Gamma(H \to \gamma\gamma)  =   \displaystyle
            \frac{\alpha^2 m_H^3}{256 \pi^3 v^2}
            \left| \sum_i y_i N_{ci} Q_i^2 F_i \right|^2,
    \eeq
with $N_{ci}$, $Q_i$ are the color factor and the electric charge
respectively for particles running in the loop. The dimensionless
loop factors for particles of spin given in the subscript are \beq
    F_1(\tau) = 2 + 3 \tau + 3\tau (2-\tau) f(\tau),\quad
    F_{1/2}(\tau) = -2\tau [1 + (1-\tau)f(\tau)],\quad
    F_0(\tau) = \tau [1 - \tau f(\tau)],
\eeq where $\tau=4m_i^2/m_H^2$ and $y_i$ is from \beq
 {\cal L}=- \frac{m_t}{v}y_t\bar{t}tH+ 2 \frac{m_{W}^2}{v} y_{_{W}} W^+ W^- H
      - 2 \frac{m_{\phi}^2}{v} y_{\phi} \phi \phi H
   .\label{rrinterii}\eeq
\begin{equation}
    f(\tau) = \left\{ \begin{array}{lr}
        [\sin^{-1}(1/\sqrt{\tau})]^2, & \tau \geq 1 \\
        -\frac{1}{4} [\ln(\eta_+/\eta_-) - i \pi]^2, & \, \tau < 1
        \end{array}  \right.\label{hggf12}
\end{equation}

\subsection{results and discussions}
\begin{figure}[tb]
%\begin{center}
 \epsfig{file=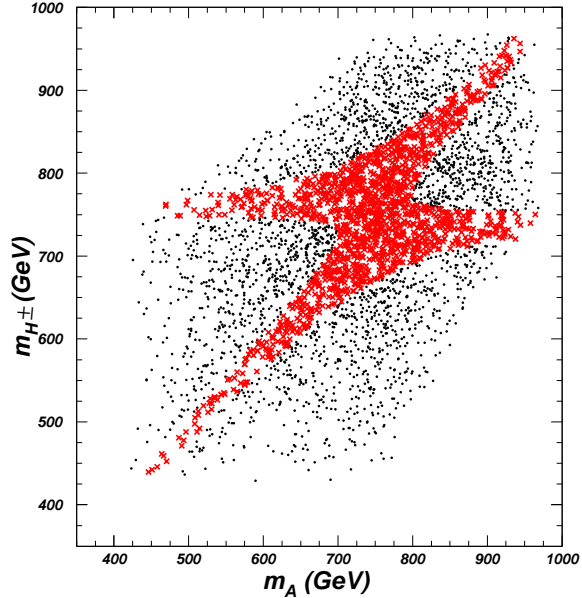,height=8.0cm}
%\end{center}
\vspace{-0.3cm} \caption{The scatter plots of surviving samples projected
 on the planes of $m_{A}$ versus $m_{H^{\pm}}$. All the samples are allowed by
the theoretical constraints. The crosses (red) and bullets (black) are respectively
allowed and excluded by the oblique parameters and $\Delta\rho$.} \label{th-ew}
\end{figure}

First we examine the allowed mass range of pseudoscalar and charged
Higgs for $m_H=750$ GeV after imposing the theoretical constraints,
oblique parameters and $\Delta\rho$, where we scan $m_A$ in the
range of $375 {\rm\  GeV} \leq m_{A}  \leq 2000  {\rm\  GeV}$. In
Fig. \ref{th-ew}, we project the surviving samples on the plane of
$m_A$ versus $m_{H^{\pm}}$. $m_A$ and $m_{H^{\pm}}$ are favored in
the range of 700 GeV and 800 GeV. In addition, the pseudoscalar and
charged Higgs masses are allowed to have sizable deviations from 750
GeV for the small mass splitting between them. Also the pseudoscalar
mass is allowed to have sizable deviation from 750 GeV for
$m_{H^{\pm}}$ is around 750 GeV.

%%%%%%%%%%%%%%%%%%%%%
%%%%%%%%%%%%%%%%%%%%

Now we calculate $R_{\gamma\gamma}$ taking $m_A\simeq m_H=750$ GeV.
Fig. \ref{th-ew} shows that $m_{H^{\pm}}$ is required to be larger
than 650 GeV for $m_A\simeq m_H$ = 750 GeV. For the decay
$H\to\gamma\gamma$, the form factor of scalar-loop is generally much
smaller than that of fermion-loop. Further, $4m^2_{H^{\pm}}/m_H^2$
has sizable deviation from 1 where the peak of form factor appears.
Therefore, the contribution of the charged Higgs to the decay
$H\to\gamma\gamma$ is much smaller than that of top quark unless the
top quark Yukawa coupling is sizably suppressed by a large
$\tan\beta$. The $W$ boson does not give the contribution to the
decay $H\to \gamma\gamma$ since the $H$ couplings to gauge bosons
are zero for $\cos(\beta-\alpha)=0$. The decays $H\to
H^{+}H^{-},~AA,~ AZ$ are kinematically forbidden, and the widths of
$H\to ~WW,~ZZ,~b\bar{b},~\tau\bar{\tau}$ are zero due to
$\cos(\beta-\alpha)=0$ and $\kappa_d=\kappa_\ell=0$. Also the $H$ coupling to
$hh$ is zero for $\cos(\beta-\alpha)=0$, and we will give the detailed explanation in
the Appendix A. The width of $H\to
H^{\pm}W^{\mp}$ can be comparable to $H\to t\bar{t}$ for the small
charged Higgs mass. However, $m_{H^{\pm}}$ is required to be larger
than 650 GeV, which leads the width to be sizably suppressed by the
large phase space. Therefore, the decay $H\to t\bar{t}$ dominates
the total width of the heavy Higgs.

In this paper we focus on the CP-conserving case where all the couplings constants of Higgs potential are taken to be real. Therefore, the
pseudoscalar $A$ coupling to $hh$ is zero. The pseudoscalar has no
decays $A\to WW,~ZZ,~hh$, and the widths of $A\to
hZ,~b\bar{b},~\tau\bar{\tau}$ are zero due to $\cos(\beta-\alpha)=0$
and $\kappa_d=\kappa_\ell=0$. Therefore, $A\to t\bar{t}$ is the
dominant decay channel. Since the charged Higgs and gauge boson do
not give the contributions to $H\to\gamma\gamma$, the top quark
plays the dominant contributions to $A\to \gamma\gamma$. In our
calculations, we use $\textsf{2HDMC}$ to calculate the total widths
of $H$ and $A$ including various possible decay channels.

%%%%%%%%%%%%%%%%%%%%%
\begin{figure}[tb]
%\begin{center}
 \epsfig{file=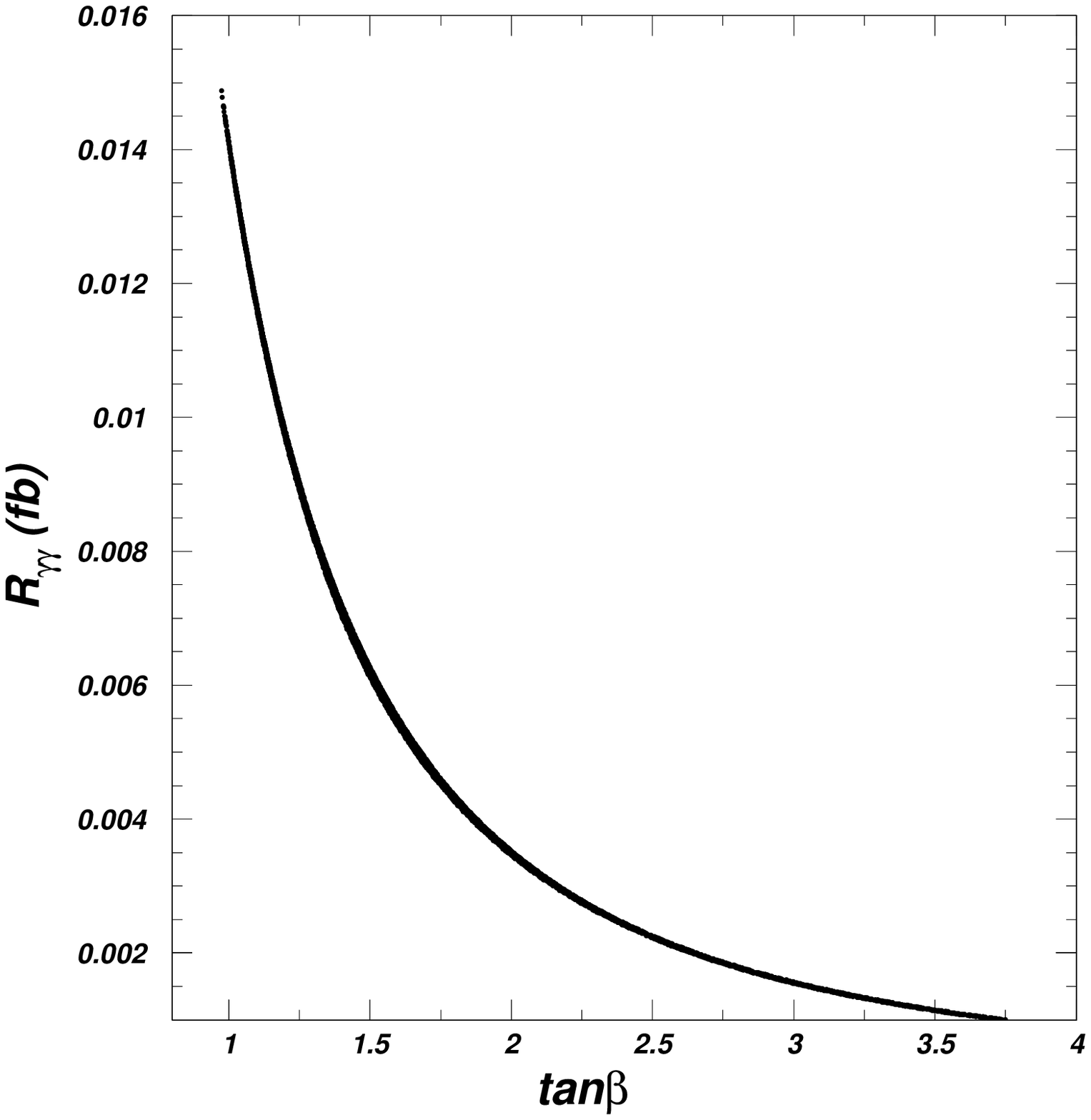,height=7.6cm}
 \epsfig{file=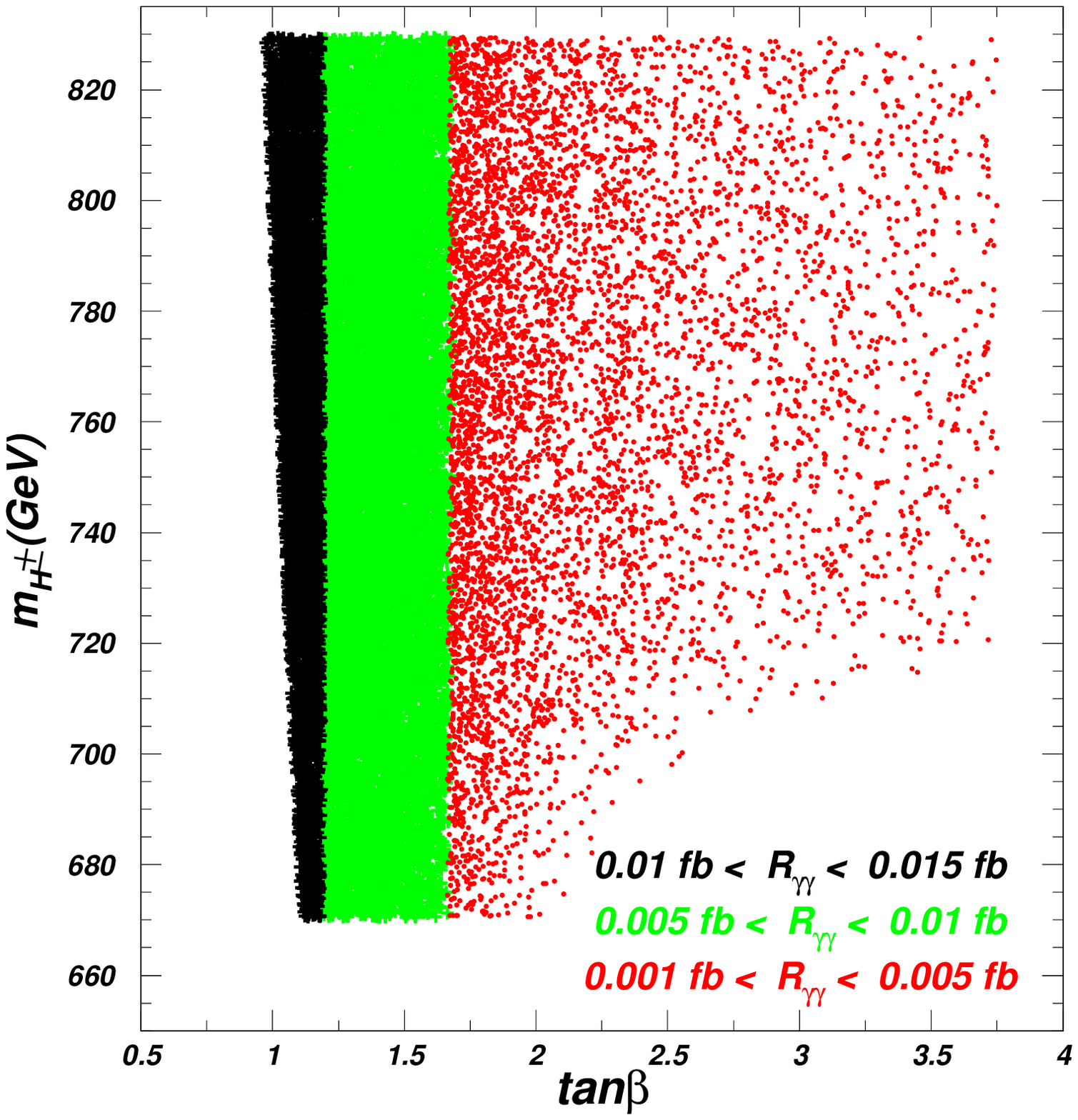,height=7.5cm}
%\end{center}
\vspace{-0.45cm} \caption{The scatter plots of surviving samples projected on the planes of $R_{\gamma\gamma}$
versus $\tan\beta$ and $m_{H^{\pm}}$
versus $\tan\beta$.} \label{rr-2h}
\end{figure}
%%%%%%%%%%%%%%%%%%%%

In Fig. \ref{rr-2h}, we project the surviving samples on the planes
of $R_{\gamma\gamma}$ versus $\tan\beta$ and $m_{H^{\pm}}$ versus
$\tan\beta$. The left panel shows the $R_{\gamma\gamma}$ increases
with decreasing of $\tan\beta$ since the $\sigma(gg\to H)$ and
$\sigma(gg\to A)$ are proportional to $1/\tan^2\beta$, and the
dependence $Br(H\to\gamma\gamma)$ and $Br(A\to\gamma\gamma)$ on
$\tan\beta$ can be canceled to some extent by the widths of
$t\bar{t}$ and $\gamma\gamma$ channels. However, due to the
constraints of $R_b$, $B\to X_s\gamma$ and $\Delta m_{B_s}$,
$\tan\beta$ is favored to be larger than 1, which leads that the
maximal value of $R_{\gamma\gamma}$ is 0.015 fb. The correlations
among $R_{\gamma\gamma}$, $\tan\beta$ and $m_{H^{\pm}}$ are shown in
the right panel.

\section{extending 2HDM with additional Higgs field}
In the minimal version of 2HDM, the production rate of 750 GeV
diphoton resonance only can reach 0.015 fb, which is smaller than
cross section observed by CMS collaboration by two order of
magnitude. Ref. \cite{1512.04921} introduces additional vector-like
quark and lepton to 2HDM in order to enhance the production rate.
Here we will extend 2HDM with additional Higgs fields and discuss
two different extensions, an inert complex Higgs triplet and a real
scalar septuplet.

\subsection{2HDM with an inert complex Higgs triplet (2HDM-IHT)}
The extension SM with a complex Higgs triplet is proposed in
\cite{htm}, called type-II seesaw model, and ref. \cite{ihtm} also
extends SM with an inert complex Higgs triplet. Here we add an inert
complex $\rm{SU(2)_L}$ triplet scalar field $\Delta$ with $Y = 2$ to
the 2HDM imposing an $Z_2$ symmetry in which the triplet is assigned
to be odd and the others even. The VEV of triplet scalar field is
zero to keep the $Z_2$ symmetry unbroken. The potential of triplet
field is written as \bea\label{m3potent}
V&=&M^2Tr(\Delta^{\dagger}{\Delta}) +
\lambda_1Tr(\Delta^{\dagger}{\Delta})^2
+\lambda_2(Tr\Delta^{\dagger}{\Delta})^2
 +
\lambda_3{\Phi^\dagger_1\Delta\Delta^{\dagger}\Phi_1}+
\lambda_4(\Phi^\dagger_1{\Phi_1})Tr(\Delta^{\dagger}{\Delta})\nonumber\\
&&+ \lambda'_3{\Phi^\dagger_2\Delta\Delta^{\dagger}\Phi_2}+
\lambda'_4(\Phi^\dagger_2{\Phi_2})Tr(\Delta^{\dagger}{\Delta}), \eea
where
\begin{eqnarray}
\Delta &=\left(
\begin{array}{cc}
\delta^+/\sqrt{2} & \delta^{++} \\
(\delta^0_r + i\delta^0_i)/\sqrt{2} & -\delta^+/\sqrt{2}\\
\end{array}
\right).
\end{eqnarray}

After the two Higgs doublet $\Phi_1$ and $\Phi_2$ acquire the VEVs, the last four terms
in Eq. (\ref{m3potent}) will give the additional contributions to the masses of components
in the triplet, respectively. At the tree-level, the Higgs triplet masses are given as,
\begin{eqnarray}\label{m3mass}
&&m^2_{\delta^{\pm\pm}}=M^2+\frac{1}{2}v^2(\lambda_4c^2_\beta+\lambda'_4s^2_\beta)\nonumber\\
&&m^2_{\delta^{\pm}}=M^2+\frac{1}{2}v^2(\lambda_4c^2_\beta+\lambda'_4s^2_\beta)
+\frac{1}{4}v^2(\lambda_3c^2_\beta+\lambda'_3s^2_\beta)\nonumber\\
&&m^2_{\delta^0_r}=m^2_{\delta^0_i}=M^2+\frac{1}{2}v^2(\lambda_4c^2_\beta+\lambda'_4s^2_\beta)
+\frac{1}{2}v^2(\lambda_3c^2_\beta+\lambda'_3s^2_\beta).
\end{eqnarray}
The charged Higgs triplet scalars couplings to the $h$ and $H$ are
given as,
\begin{eqnarray}\label{m3coupling}
&&h\delta^{+}\delta^{-}:~-\frac{1}{2}v\left(\lambda_3'c_\alpha s_\beta-\lambda_3s_\alpha c_\beta
+2\lambda_4'c_\alpha s_\beta-2\lambda_4s_\alpha c_\beta\right)\nonumber\\
&&H\delta^{+}\delta^{-}:~-\frac{1}{2}v\left(\lambda_3's_\alpha s_\beta+\lambda_3c_\alpha c_\beta
+2\lambda_4's_\alpha s_\beta+2\lambda_4c_\alpha c_\beta\right)\nonumber\\
&&h\delta^{++}\delta^{--}:~-v\left(\lambda_4'c_\alpha s_\beta-\lambda_4s_\alpha c_\beta\right)\nonumber\\
&&H\delta^{++}\delta^{--}:~-v\left(\lambda_4's_\alpha s_\beta+\lambda_4c_\alpha c_\beta\right).
%h\delta^{0}_r\delta^{0}_r=h\delta^{0}_i\delta^{0}_i:~
%-v\left(\lambda_3'c_\alpha s_\beta-\lambda_3s_\alpha c_\beta
%+\lambda'_4c_\alpha s_\beta-\lambda_4s_\alpha c_\beta\right)\nonumber\\
%&&H\delta^{0}_r\delta^{0}_r=H\delta^{0}_i\delta^{0}_i:~
%-v\left(\lambda_3's_\alpha s_\beta+\lambda_3c_\alpha c_\beta
%+\lambda'_4s_\alpha s_\beta+\lambda_4c_\alpha c_\beta\right)
\end{eqnarray}

\subsection{2HDM with a real scalar septuplet (2HDM-RSS)}
The extension of SM with the complex and real scalar septuplet has
been studied in \cite{sepcomplex} and \cite{sepreal}. Here we
introduce a real scalar septuplet to the 2HDM assuming that the
septuplet does not develop the VEV. The potential of the septuplet
field $\Sigma$ is written as \bea\label{m7potent}
V&=&M^2\Sigma^{\dagger}\Sigma + \lambda_1(\Sigma^{\dagger}\Sigma)^2
 + \frac{\lambda_2}{48}(\Sigma^{\dagger}T^aT^b\Sigma)^2\nonumber\\
&&+ \lambda_3(\Phi^\dagger_1{\Phi_1})(\Sigma^{\dagger}\Sigma) +
\lambda'_3(\Phi^\dagger_2{\Phi_2})(\Sigma^{\dagger}\Sigma), \eea
where \beq
\Sigma=\frac{1}{\sqrt{2}}\left(T^{+++},~T^{++},~T^+,~T^0,~T^-,~T^{--},~T^{---}\right)^T.
\eeq

After the two Higgs doublet $\Phi_1$ and $\Phi_2$ develop the VEVs,
the last two terms in Eq. (\ref{m7potent}) will give the additional
contributions to the masses of all components of $\Sigma$. The
septuplet scalars are degenerate at the tree-level and their mass
are \beq m^2_{\Sigma} = M^2 + \frac{1}{2}v^2(\lambda_3c^2_\beta +
\lambda'_3s^2_\beta). \eeq

The charged components of septuplet couplings to the $h$ and $H$
are,
\begin{eqnarray}
&&hT^{+}T^{-}=hT^{++}T^{--}=hT^{+++}T^{---}:~-v\left(\lambda_3'c_\alpha s_\beta-\lambda_3s_\alpha c_\beta\right)\nonumber\\
&&HT^{+}T^{-}=HT^{++}T^{--}=HT^{+++}T^{---}:~-v\left(\lambda_3's_\alpha s_\beta+\lambda_3c_\alpha c_\beta\right).
\end{eqnarray}

\subsection{calculations and discussions}
In the original 2HDM, the production rate for the 750 GeV diphoton
resonance increases with decreasing of $\tan\beta$. However, the
$R_b$ and $B$ flavor observables disfavor $\tan\beta$ to be smaller
than 1. Therefore, in the following calculations we will take \beq
\tan\beta=1,~~~\sin(\beta-\alpha)=1. \label{sba=1}\eeq Further, in order to
forbid the new charged Higgses altering the 125 GeV Higgs decay into
$\gamma\gamma$ via one-loop effects, we require the light CP-even
Higgs couplings to these charged Higgses to be zero, which leads
for Eq. (\ref{sba=1}),
\beq\lambda_4=-\lambda'_4~ {\rm and}~ \lambda_3=-\lambda'_3 {\rm
~for~ 2HDM-IHT,}\label{m3h=0}\eeq \beq\lambda_3=-\lambda'_3 ~{\rm for~ 2HDM}-{\rm
RSS.}\eeq For Eq. (\ref{sba=1}) and Eq. (\ref{m3h=0}), the triplet scalar masses in the 2HDM-IHT
become degenerate, \beq
m^2_{\delta^{\pm\pm}}=m^2_{\delta^{\pm}}=m^2_{\delta^{0}_r}=m^2_{\delta^{0}_i}=M^2.
\eeq The $H$ couplings to the charged components of triplet are,
\beq\label{m3coupling2}
H\delta^{+}\delta^{-}:-\left(\lambda_4+\frac{1}{2}\lambda_3\right)v,~~~~~H\delta^{++}\delta^{--}:-\lambda_4v.
\eeq Similarly, the septuplet scalar masses of the 2HDM-RSS are \beq
m_{\Sigma} = M. \eeq The $H$ couplings to the charged components of
septuplet are, \beq
HT^{+}T^{-}=HT^{++}T^{--}=HT^{+++}T^{---}:~-\lambda_3v. \eeq

For the Higgs triplet and septuplet, the mass splitting among the
components can be induced by loop corrections, and the charged
components are very slightly heavier than the neutral components
\cite{ihtm,sepreal}. These mass splittings are negligibly small, and
the two extensions can be well consistent with the oblique
parameters. Both the Higgs triplet and septuplet have no
 interactions with fermions, and the interactions with
gauge bosons have to contain two components simultaneously, which
makes them to be hardly constrained by the low energy observables
and collider experimental searches. For the inert scalars, the multi-lepton +  $\eslash_T$ is
regarded as one of the most promising channels at the LHC. Taking the inert Higgs triplet model as
an example, the charged scalars can be produced at the LHC through Drell-Yan process,
\beq
q\bar{q}\to \delta^+ \delta^-,~\delta^{++}\delta^{--},~\delta^{\pm}\delta^{\mp\mp},
\eeq
and the charged scalars can have the following cascade decay assuming
 $m_{\delta^{\pm\pm}}>m_{\delta^{\pm}}>m_{\delta^0_i}>m_{\delta^0_r}$ ($\delta^0_r$ is the stable particle),
\bea
&&\delta^{\pm\pm}\to \delta^{\pm}~W^{\pm(*)}~ (W^{\pm(*)}\to l^\pm~ \nu), \nonumber\\
&&\delta^{\pm}\to W^{\pm(*)}~\delta^0_r \to  l^\pm~ \nu~ \delta^0_r,\nonumber\\
&&\delta^{\pm}\to W^{\pm(*)}~ \delta^0_i \to l^\pm~ \nu ~\delta^0_r ~Z^{(*)} ~(Z^{(*)} > l^\pm l^\mp).
\eea

The ATLAS and CMS collaborations have searched the $2l+\eslash_T$ \cite{atlas-1403.5294,cms-1405.7570},
 $3l+\eslash_T$ \cite{cms-1405.7570,atlas-1402.7029} and $4l+\eslash_T$ \cite{cms-1405.7570},
 and set the limits on the next-to-lightest neutralino, the lightest-neutralino and chargino
 in the supersymmetric model. The lower bound of their masses can be up to hundreds of GeV
for the large mass splittings. The ATLAS and CMS searches for the
multi-lepton + $\eslash_T$ signals rely on triggers that require
$p_T>$ 20 GeV for the transverse momentum of one lepton at least.
The produced leptons tend to become soft with the decreasing of the
mass splittings, and the soft leptons are difficult to be detected
due to the lepton $p_T$ requirements of the search. For example,
using the $3l+\eslash_T$ signal at the 14 TeV LHC, the 300 fb$^{-1}$
of data is required to discover these supersymmetric spectra with
dark mass between 40 GeV and 140 GeV for the mass splittings drop
down to 9 GeV \cite{1602.00590}. In this paper, the charged and
neutral components of the inert scalar multiplets are degenerate at
the tree-level, and the loop corrections only make the charged
components to be very slightly heavier than the neutral components
\cite{ihtm,sepreal}. For such small mass splitting, the leptons of
the multi-lepton + $\eslash_T$ event are very soft. Therefore, the
inert scalars are free from the constraints of the ATLAS and CMS
searches for the multi-lepton + $\eslash_T$ at the 8 TeV LHC,
 and even difficult to be detected at the 14 TeV LHC with the more high integrated luminosity. Note that in order to
enhance the 750 GeV Higgs decay into diphoton, in this paper we take
the masses of the inert scalars to be in the range of 375 GeV and
500 GeV. Such range of mass will lead the relic density to be much
smaller than the observed value although the lightest neutral
component is stable \cite{ihtm}. Therefore, to produce the observed
relic density \cite{planck}, some other dark matter sources need be
introduced, which is beyond the scope of this paper.

For $\tan\beta=1$ and $\sin(\beta-\alpha)=1$, the $H$ couplings to the charged Higgs $H^{\pm}$ of the original 2HDM
is zero. Therefore, the $H^{\pm}$ does not give the contributions to the decay $H\to \gamma\gamma$. For 2HDM-IHT, the doubly charged
 and singly charged Higgses $\delta^{\pm\pm}$ and $\delta^{\pm}$ give additional contributions to the decay $H\to \gamma\gamma$,
which are sensitive to the mass $M$ and the coupling constants
$\lambda_4$, $\lambda_3$. The degenerate masses of the triplet
scalars are taken to be larger than 375 GeV, which makes the 750 GeV
Higgs decays into triplet scalars to be kinematically forbidden. The
perturbativity requires the absolute values of $\lambda_4$ and
$\lambda_3$ in the quartic terms to be smaller than $4\pi$. The
stability of the potential favors $\lambda_4$ and $\lambda_3$ to be
larger than 0, and gives the lower bound of $\lambda_4$ and
$\lambda_3$ for they are smaller than zero \cite{htmtheo}. Thus we
take 0 $<\lambda_4<$ $4\pi$ and fix $\lambda_3=3$ \cite{1112.5453}.
Because $\delta^{\pm\pm}$ has an electric charge of $\pm 2$, the
$\delta^{\pm\pm}$ contributions are enhanced by a relative factor 4
in the amplitude of $H\to \gamma\gamma$, see Eq. (\ref{hrrgs}),
which can help $\delta^{\pm\pm}$ contributions dominate over the
other particle contributions. Since there are the same sign between
$H\delta^{++}\delta^{--}$ and $H\delta^{+}\delta^{-}$, the
$\delta^{\pm\pm}$ and $\delta^{\pm}$ contributions are constructive
each other.

%%%%%%%%%%%%%%%%%%%%%
\begin{figure}[tb]
%\begin{center}
 \epsfig{file=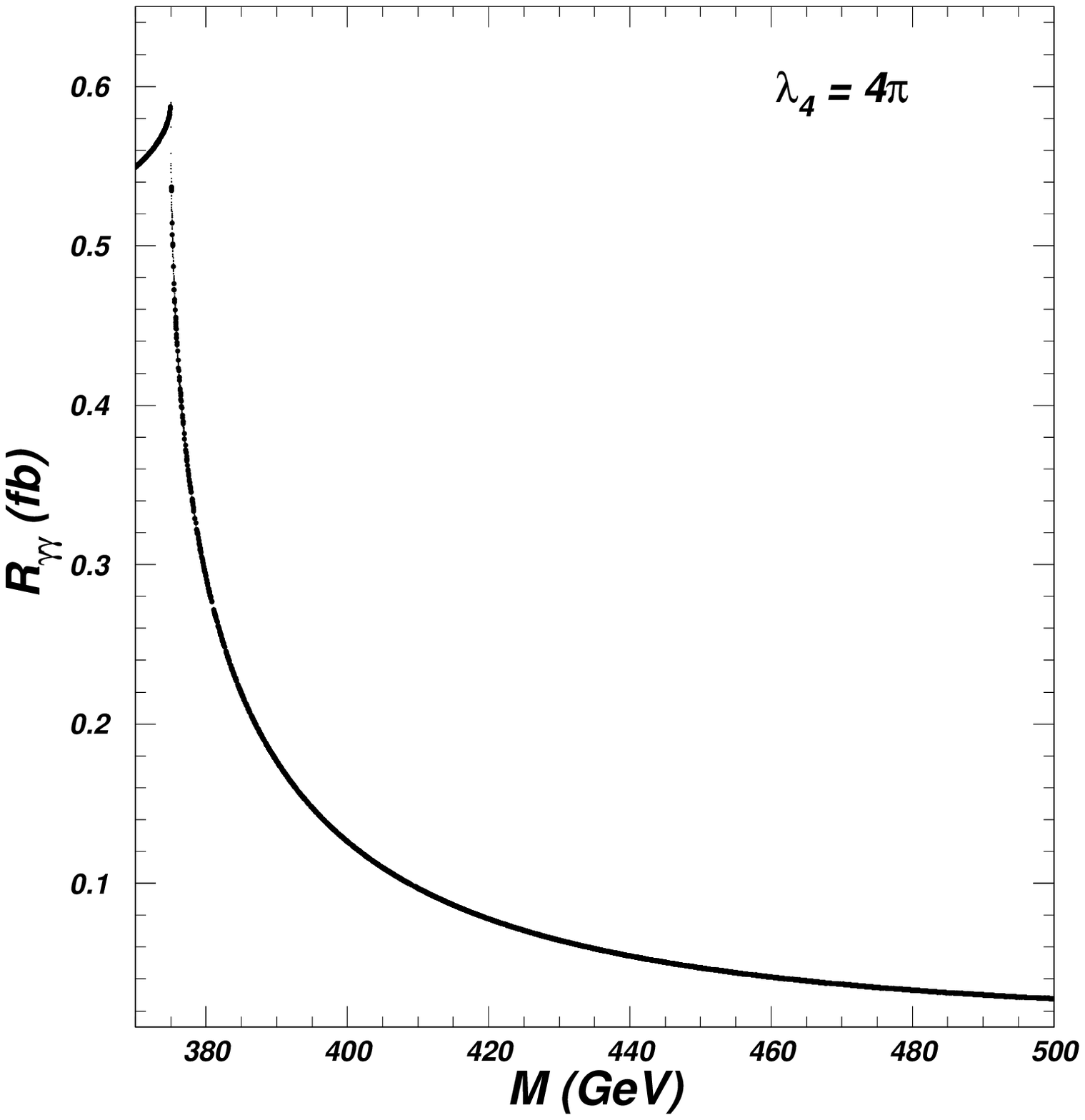,height=5.7cm}
  \epsfig{file=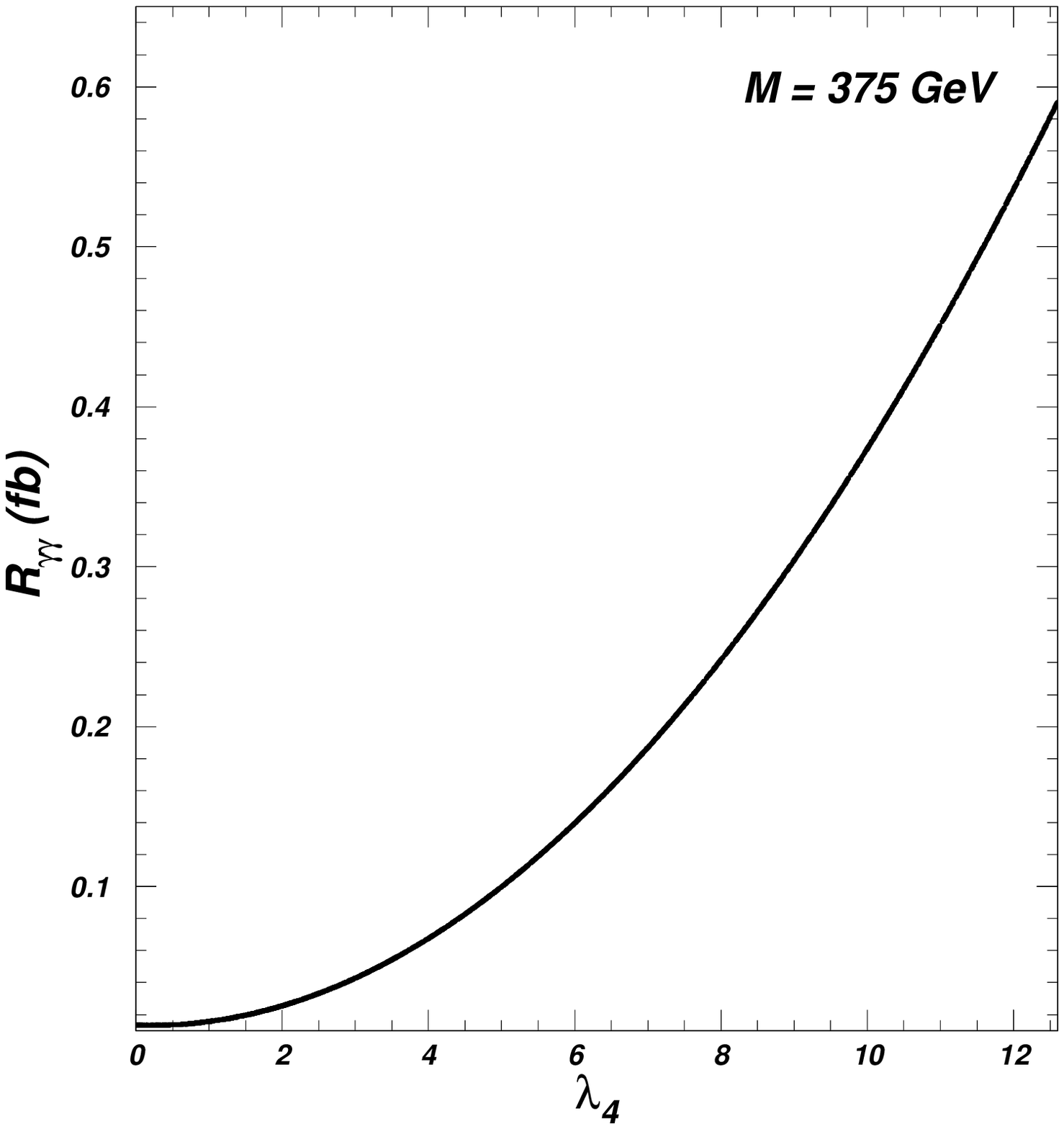,height=5.7cm}
 \epsfig{file=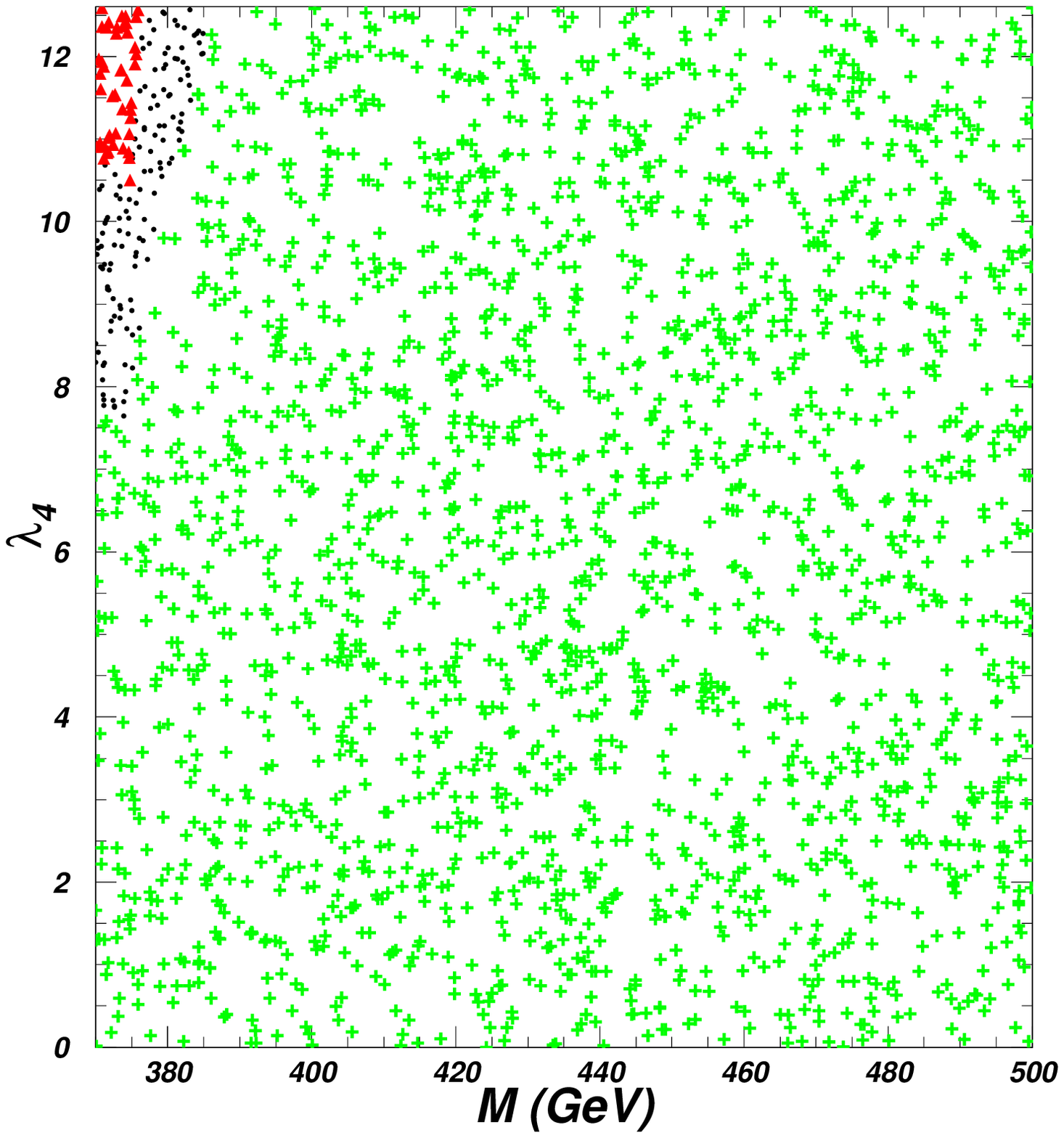,height=5.7cm}
 %\end{center}
\vspace{-1.25cm} \caption{$R_{\gamma\gamma}$ versus $M$, $R_{\gamma\gamma}$ versus $\lambda_4$
and $\lambda_4$ versus $M$ in the 2HDM with an inert complex Higgs triplet. In the right panel
 $R_{\gamma\gamma}<$ 0.2 fb for the pluses (green), 0.2 fb $<R_{\gamma\gamma}<$ 0.4 fb for the bullets (black) and
0.4 fb $<R_{\gamma\gamma}<$ 0.6 fb for the triangles (red).} \label{rr-2hm3}
\end{figure}
%%%%%%%%%%%%%%%%%%%%

%%%%%%%%%%%%%%%%%%%%%
\begin{figure}[tb]
%\begin{center}
 \epsfig{file=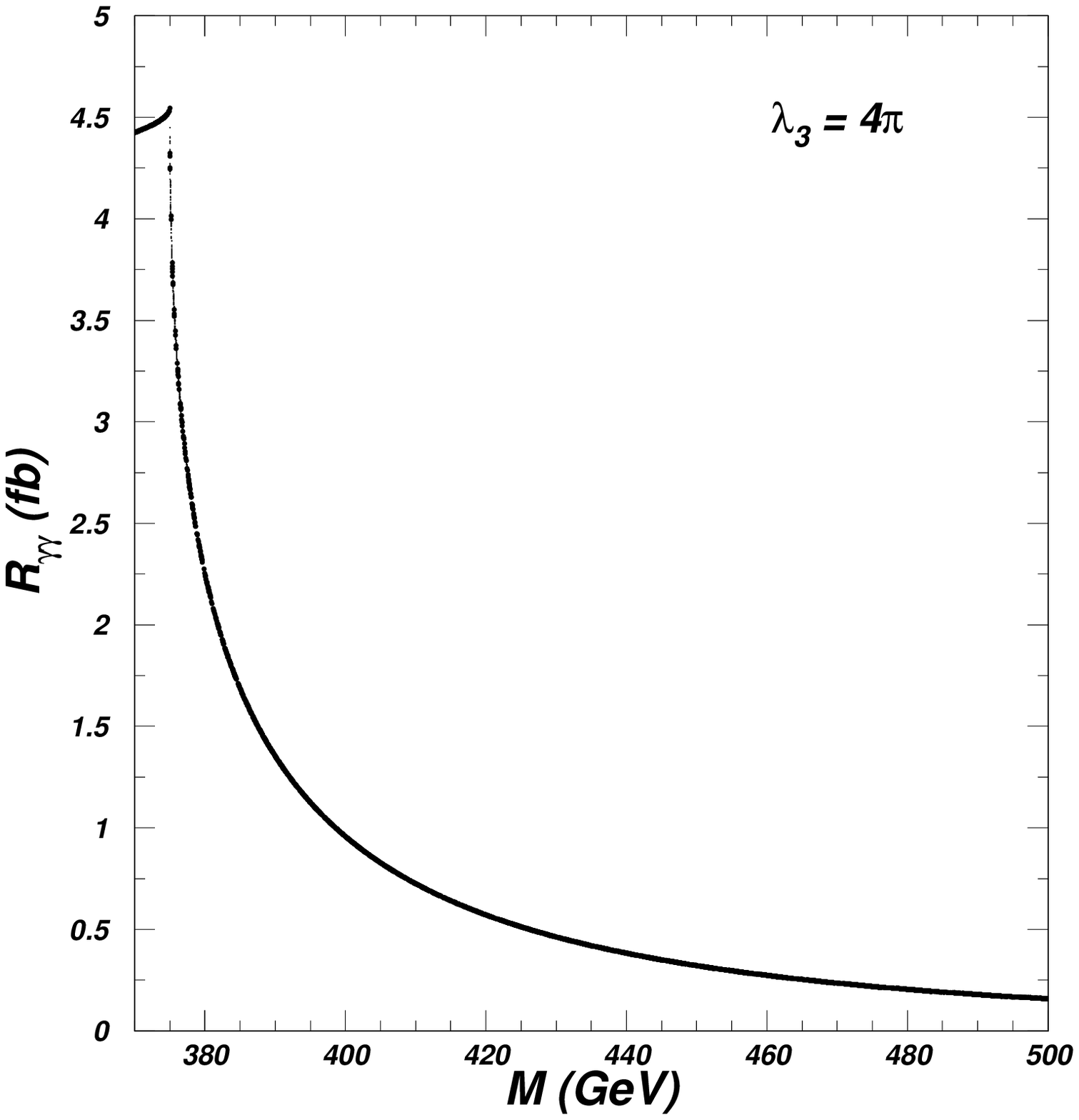,height=5.7cm}
  \epsfig{file=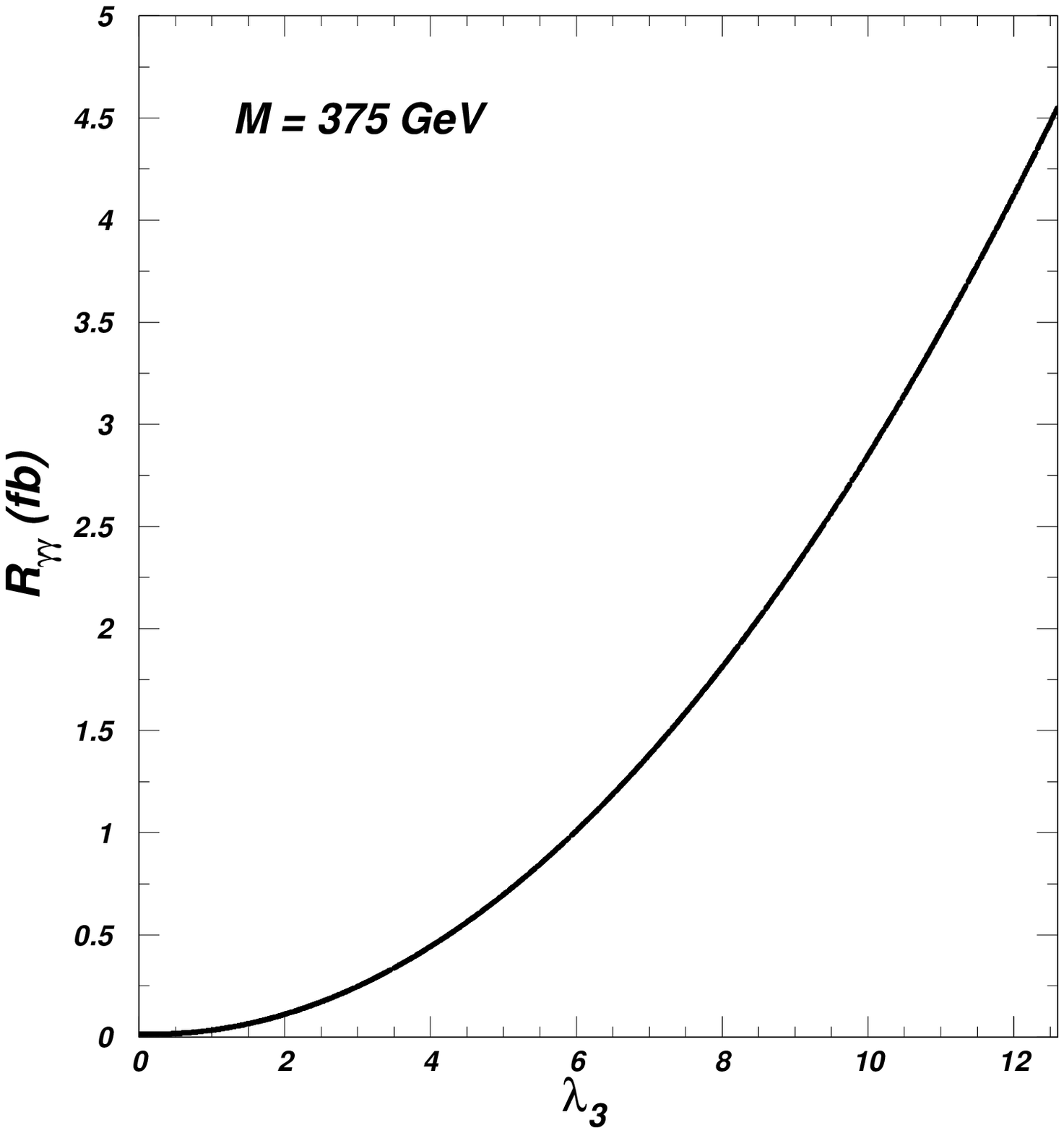,height=5.7cm}
 \epsfig{file=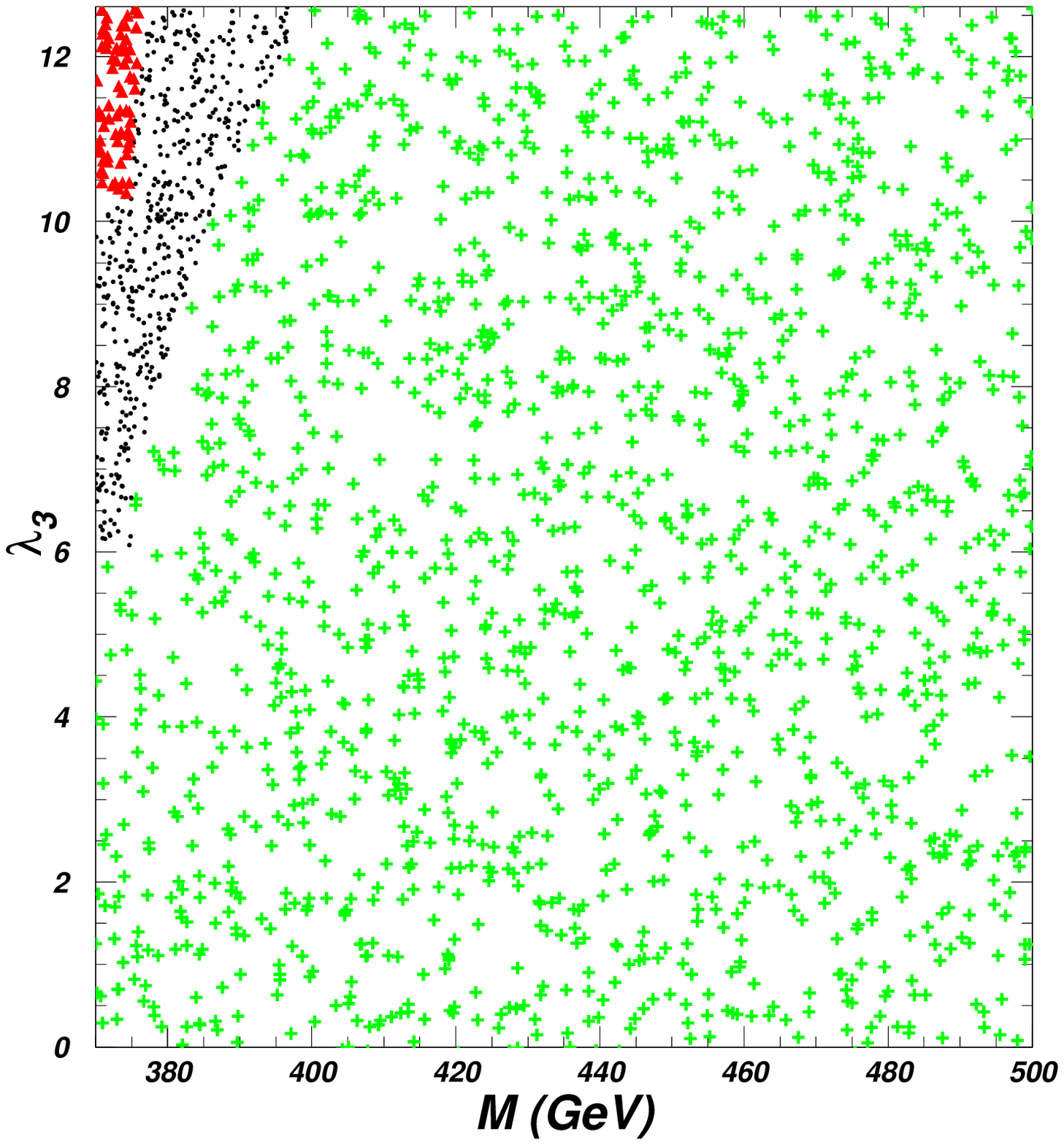,height=5.7cm}
 %\end{center}
\vspace{-1.25cm} \caption{$R_{\gamma\gamma}$ versus $M$, $R_{\gamma\gamma}$ versus $\lambda_3$
and $\lambda_3$ versus $M$ in the 2HDM with a real scalar septuplet. In the right panel
 $R_{\gamma\gamma}<$ 1.0 fb for the pluses (green), 1.0 fb $<R_{\gamma\gamma}<$ 3.0 fb for the bullets (black) and
3.0 fb $<R_{\gamma\gamma}<$ 4.6 fb for the triangles (red).} \label{rr-2hm7}
\end{figure}
%%%%%%%%%%%%%%%%%%%%

In the Fig. \ref{rr-2hm3}, we project the samples of 2HDM-IHT on the
planes of $R_{\gamma\gamma}$ versus $M$, $R_{\gamma\gamma}$ versus
$\lambda_4$ and $\lambda_4$ versus $M$. The left panel shows that
$R_{\gamma\gamma}$ has a peak around $M=$ 375 GeV, and decreases
rapidly with increasing of $M$. The characteristic is determined by
the form factor $F_0(\tau)$ in the $H\to\gamma\gamma$.
$R_{\gamma\gamma}$ can reach 0.6 fb for $M=$ 375 GeV and $\lambda_4=
{4\pi}$, but is still much smaller than the cross section for the
750 GeV diphoton resonance observed by CMS and ATLAS collaborations.

For the 2HDM-RSS, $T^{\pm\pm\pm}$, $T^{\pm\pm}$ and $T^{\pm}$ give
additional contributions to the decay $H\to \gamma\gamma$, which are
sensitive to the mass $M$ and the coupling $\lambda_3$. The
perturbativity requires the absolute value of $\lambda_3$ to be
smaller than $4\pi$, and $\lambda_3 >$ 0 is free from the
constraints of the potential stability \cite{sepreal}. Compared to
the 2HDM-IHT, $T^{\pm\pm\pm}$ of 2HDM-RSS has an electric charge of
$\pm 3$, and the $T^{\pm\pm\pm}$ contributions are enhanced by a
relative factor 9 in the amplitude of $H\to \gamma\gamma$, which
makes $T^{\pm\pm\pm}$ contributions to dominate over the other
particle contributions. Further, since there are the same sign
between $HT^{+++}T^{---}$,
 $HT^{++}T^{--}$ and $HT^{+}T^{-}$, their contributions are constructive each other. Therefore,
the width of $H\to \gamma\gamma$ of 2HDM-RSS can be much larger than
that of 2HDM-ITH, approximate 8 times for the same Higgs coupling
and mass. In the Fig. \ref{rr-2hm7}, we project the samples of
2HDM-RSS on the planes of $R_{\gamma\gamma}$ versus $M$,
$R_{\gamma\gamma}$ versus $\lambda_3$ and $\lambda_3$ versus $M$.
$R_{\gamma\gamma}$ can reach 4.6 fb for $M=$ 375 GeV and $\lambda_3=
{4\pi}$, which is approximate 8 times of the maximal value of of
2HDM-IHT. The right panel shows that $R_{\gamma\gamma}>$ 1 fb favors
$M<$ 400 GeV and $\lambda_3>$ 6, and $R_{\gamma\gamma}>$ 3 fb for
$M<$ 380 GeV and $\lambda_3>$ 10. Therefore, 2HDM-RSS can
accommodate the 750 GeV diphoton resonance observed by the CMS and
ATLAS at the LHC.

\section{Conclusion}
In this paper, we first consider various theoretical and
experimental constraints, and examine the implications of the 750
GeV diphoton resonance on the two-Higgs-doublet model. We find
the pseudoscalar and charged Higgs masses are favored in the range of
700 GeV and 800 GeV, and their masses are allowed to have sizable
deviations from 750 GeV for the small mass splitting between them.
Also the pseudoscalar mass is allowed to have sizable deviation from
750 GeV for the charged Higgs mass around 750 GeV. In the
two-Higgs-doublet model, the production rate for 750 GeV diphoton
resonance is smaller than the cross section observed at LHC by two
order magnitude. In order to accommodate the 750 GeV diphoton resonance, we
respectively introduce an inert complex Higgs triplet and a real
scalar septuplet to the two-Higgs-doublet model. The multi-charged
scalars in these models can enhance the branching ratio of $H\to
\gamma\gamma$ sizably. The production rate for the 750 GeV diphoton
resonance can be enhanced to 0.6 fb for 2HDM with an inert Higgs
triplet and 4.5 fb for 2HDM with a real scalar septuplet. The latter
can give a valid explanation for
 the 750 GeV diphoton resonance at the LHC.

\section*{Acknowledgment}
This work has been supported in
part by the National
Natural Science Foundation of China under grant No. 11575152, and by the Spanish
Government and ERDF funds from the EU Commission
[Grants No. FPA2014-53631-C2-1-P, SEV-2014-0398, FPA2011-23778].

\appendix
\section{The coupling of $Hhh$}
The scalar potential shown in the Eq. (\ref{V2HDM}) is expressed in
the physical basis where both $\Phi_1$ and $\Phi_2$ have the
non-zero VEVs. It is more convenient to understand the coupling
$Hhh$ in the Higgs basis where the two scalar doublets are given as\beq
\label{cphiggsbasisfields} H_1=\begin{pmatrix}G^+\\
\frac{1}{\sqrt{2}}\,(v+\rho_1+iG_0)\end{pmatrix}\equiv
\Phi_1\cb+\Phi_2\sb\,, ~~ H_2=\begin{pmatrix} H^+\\
\frac{1}{\sqrt{2}}\,(\rho_2+iA)\end{pmatrix}\equiv
-\Phi_1\sb+\Phi_2\cb\,. \eeq In the Higgs basis,  the $H_1$ field
has a VEV $v=$246 GeV, and the VEV of $H_2$ field is zero.

The scalar potential in the physical basis (as shown in the Eq. (\ref{V2HDM}) ) can be expressed in the Higgs basis \cite{yun},
\bea
\mathcal{V}&=& Y_1 H_1^\dagger H_1+Y_2 H_2^\dagger H_2
+Y_3[H_1^\dagger H_2+{\rm h.c.}]
+\half Z_1(H_1^\dagger H_1)^2
+\half Z_2(H_2^\dagger H_2)^2
+Z_3(H_1^\dagger H_1)(H_2^\dagger H_2)
\nonumber\\[8pt]
&&\quad\quad
+Z_4(H_1^\dagger H_2)(H_2^\dagger H_1)
+\left\{\half Z_5(H_1^\dagger H_2)^2
+\big[Z_6(H_1^\dagger H_1)
+Z_7(H_2^\dagger H_2)\big]
H_1^\dagger H_2+{\rm h.c.}\right\}, \label{potZ}
\eea
where the $Y_i$ are real linear combinations of the $m_{ij}^2$ and the $Z_i$ are real linear combinations of the $\lambda_i$.
For $\lambda_6=\lambda_7=0$, we simply have \cite{yun}
\bea
Z_1 & \equiv & \lambda_1 c^4_\beta+\lambda_2 s^4_\beta+\half\lamtil s^2_{2\beta}\,,\label{zeeone}\\
Z_2 & \equiv & \lambda_1 s^4_\beta+\lambda_2 c^4_\beta+\half\lamtil s^2_{2\beta}\,,\label{zeetwo}\\
Z_i & \equiv & \tfrac{1}{4} s^2_{2\beta}\bigl[\lambda_1+\lambda_2-2\lamtil\bigr]+\lambda_i\,,\quad \text{(for $i=3,4$ or 5)}\,,\label{zeefive}\\
Z_6 & \equiv & -\half s_{2\beta}\bigl[\lambda_1 c^2_\beta-\lambda_2 s^2_\beta-\lamtil c_{2\beta}\bigr]\,,\label{zeesix} \\
Z_7 & \equiv & -\half s_{2\beta}\bigl[\lambda_1 s^2_\beta-\lambda_2 c^2_\beta+\lamtil c_{2\beta}\bigr]\,,\label{zeeseven}
\eea
where $c_{2\beta}=\cos 2\beta$, $s_{2\beta}=\sin 2\beta$ and $\lambda_{345}=\lambda_{3}+\lambda_4 +\lambda_5$.

 The
$H^+$ and $A$ are the mass eigenstates of the charged Higgs boson and
CP-odd Higgs boson, and their masses are given by \bea
m_{H^+}^2&=&Y_2+\half Z_3 v^2\,, \nonumber\\
m_A^2&=& Y_2+\half(Z_3+Z_4-Z_5)v^2.
\eea

The physical CP-even Higgs bosons $h$ and $H$ are the linear combination of $\rho_1$ and $\rho_2$,
\bea
H &=&\rho_1 \cos(\beta-\alpha) - \rho_2 \sin(\beta-\alpha),\nonumber\\
h &=&\rho_1 \sin(\beta-\alpha) + \rho_2 \cos(\beta-\alpha).
\eea
For $\cos(\beta-\alpha)=0$, there is no mixing of $h$ and $H$, which requires $Z_6=0$ and leads to
\beq
H=- \rho_2,~~~~~~h=\rho_1.
\label{mixing0}
\eeq
For the Eq. (\ref{mixing0}), the other terms except for the $Z_6$ term in the Eq. (\ref{potZ}) do not produce
the coupling of $Hhh$. Therefore, the $H$ coupling to $hh$ is zero for $\cos(\beta-\alpha)=0$.

Note that $\cos(\beta-\alpha)$ denotes the coupling of $H$ and gauge
bosons normalized to SM Higgs. Both $\cos(\beta-\alpha)$ and the
coupling of $Hhh$ are the physical observables and 
basis-independent. Therefore, for $\cos(\beta-\alpha)=0$, the
coupling of $Hhh$ equals to zero in the physical basis and Higgs
basis. In fact, in the physical basis, the Higgs potential shown in
Eq. (\ref{V2HDM}) gives the coupling of $Hhh$ \cite{yun}, \bea
g_{Hhh}=&& -\frac{\cos(\beta-\alpha)}{v}\biggl\{4\overline{m}^{\,2}-m_H^2-2m_h^2 \nonumber\\
&&+2(3\overline{m}^{\,2}-m_H^2-2m_h^2)
[\sin(\beta-\alpha)\cot 2\beta-\cos(\beta-\alpha)]\cos(\beta-\alpha)\biggr\}\,,\label{Hhh}
\eea
with $\overline{m}^{\,2}=m_A^2 + Z_5 v^2+ \frac{1}{2} (Z_6 - Z_7) \tan 2\beta v^2$. Where $Z_5$, $Z_6$ and $Z_7$
can be expressed using the coupling constants of Higgs potential in the physical basis according to
Eq. (\ref{zeefive}), Eq. (\ref{zeesix}) and Eq. (\ref{zeeseven}). The Eq. (\ref{Hhh}) explicitly shows that
the coupling of $Hhh$ equals to zero for $\cos(\beta-\alpha)=0$ in the physical basis.

%%%%%%%%%%%%%%%%%%%%%%%%%%%%%%%%%%%%%%%%%%%%%%%%%%%%
%%%%%%%%%%%%%%%%%%%%%%%%%%%%%%%%%%%%%%%%%%%%%%%%%%%


\begin{thebibliography}{99}
\bibitem{750} ATLAS and CMS Collaborations - Dec. 15th talks by Jim Olsen and Marumi Kado, ATLAS
and CMS physics results from Run 2,
https://indico.cern.ch/event/442432/.

\bibitem{1512.04939} S. D. Chiara, L. Marzola, M. Raidal, arXiv:1512.04939.

\bibitem{dijet} V. Khachatryan et al. [CMS Collaboration], arXiv:1512.01224.
\bibitem{ditt} V. Khachatryan et al. [CMS Collaboration], arXiv:1506.03062.


\bibitem{750work1} K. Harigaya, Y. Nomura, arXiv:1512.04850; Y. Mambrini, G. Arcadi, A. Djouadi,
arXiv:1512.04913; M. Backovic, A. Mariotti, D. Redigolo, arXiv:1512.04917;
Y. Nakai, R. Sato, K. Tobioka, arXiv:1512.04924;
S. Knapen, T. Melia, M. Papucci, K. Zurek, arXiv:1512.04928;
D. Buttazzo, A. Greljo, D. Marzocca, arXiv:1512.04929;
A. Pilaftsis, arXiv:1512.04931;
R. Franceschini, G. F. Giudice, J. F. Kamenik, M. Mc-
Cullough, A. Pomarol, R. Rattazzi, M. Redi, F. Riva,
A. Strumia, R. Torre, arXiv:1512.04933;
 T. Higaki, K. S. Jeong, N. Kitajima, F. Takahashi, arXiv:1512.05295;
 S. D. McDermott, P. Meade, H. Ramani, arXiv:1512.05326;
J. Ellis, S. A. R. Ellis, J. Quevillon, V. Sanz, T. You, arXiv:1512.05327;
M. Low, A. Tesi, L.-T. Wang, arXiv:1512.05328;
B. Bellazzini, R. Franceschini, F. Sala, J. Serra, arXiv:1512.05330;
R. S. Gupta, S. Jger, Y. Kats, G. Perez, E. Stamou, arXiv:1512.05332;
C. Petersson, R. Torre, arXiv:1512.05333;
 E. Molinaro, F. Sannino, N. Vignaroli, arXiv:1512.05334.

\bibitem{750work2}B. Dutta, Y. Gao, T. Ghosh, I. Gogoladze, T. Li, arXiv:1512.05439;
Q.-H. Cao, Y. Liu, K.-P. Xie, B. Yan, D.-M. Zhang, arXiv:1512.05542;
S. Matsuzaki, K. Yamawaki, arXiv:1512.05564;
A. Kobakhidze, F. Wang, L. Wu, J. M. Yang, M. Zhang, arXiv:1512.05585; R. Martinez, F. Ochoa, C.F. Sierra, arXiv:1512.05617;
P. Cox, A. D. Medina, T. S. Ray, A. Spray, arXiv:1512.05618; J. M. No, V. Sanz, J. Setford, arXiv:1512.05700;
S. V. Demidov, D. S. Gorbunov, arXiv:1512.05723; W. Chao, R. Huo, J.-H. Yu, arXiv:1512.05738;
S. Fichet, G. v. Gersdorff, C. Royon, arXiv:1512.05751; D. Curtin, C. B. Verhaaren, arXiv:1512.05753;
L. Bian, N. Chen, D. Liu, J. Shu, arXiv:1512.05759;
J. Chakrabortty, A. Choudhury, P. Ghosh, S. Mondal, T. Srivastava, arXiv:1512.05767;
A. Ahmed, B. M. Dillon, B. Grzadkowski, J. F. Gunion, Y. Jiang, arXiv:1512.05771;
P. Agrawal, J. Fan, B. Heidenreich, M. Reece, M. Strassler, arXiv:1512.05775;
C. Csaki, J. Hubisz, J. Terning,  arXiv:1512.05776;
A. Falkowski, O. Slone, T. Volansky,  arXiv:1512.05777;
D. Aloni, K. Blum, A. Dery, A. Efrati, Y. Nir, arXiv:1512.05778;
Y. Bai, J. Berger, R. Lu, arXiv:1512.05779;
F. Wang, L. Wu, J. M. Yang, M. Zhang, arXiv:1512.06715;
J. Cao, C. Han, L. Shang, W. Su, J. M. Yang, Y. Zhang, arXiv:1512.06728;
J. J. Heckman, arXiv:1512.06773;
J. Chang, K. Cheung, C.-T. Lu, arXiv:1512.06671;
D. Becirevic, E. Bertuzzo, O. Sumensari, R. Z. Funchal, arXiv:1512.05623;
A. Alves, A. G. Dias, K. Sinha, arXiv:1512.06091.


\bibitem{1512.04921} A. Angelescu, A. Djouadi, G. Moreau, arXiv:1512.04921.

\bibitem{2h-poten} R. A. Battye, G. D. Brawn, A. Pilaftsis, \JHEP1108, 020 (2011).

\bibitem{a2hm} A. Pich, P. Tuzon, \PRD80, 091702 (2009).

\bibitem{a2hm2} W. Altmannshofer, S. Gori and G. D. Kribs, \PRD86, 115009 (2012).

\bibitem{a2hfree} V. Barger, L. L. Everett, H. E. Logan, G. Shaughnessy, \PRD88, 115003 (2013).

\bibitem{2hc-1} D. Eriksson, J. Rathsman, O. St{\aa}l, \CPC181, 189-205
(2010); \CPC181, 833-834 (2010).

\bibitem{2hpert} B. Grinstein, C. W. Murphy, P. Uttayarat, arXiv:1512.04567.

\bibitem{hb-1} P. Bechtle, O. Brein, S. Heinemeyer, G. Weiglein, K. E.
Williams, \CPC181, 138-167 (2010).
\bibitem{hb-2} P. Bechtle, O. Brein, S. Heinemeyer, O. St{\aa}l, T.
Stefaniak, G. Weiglein, K. E. Williams, \EPJC74, 2693 (2014).

\bibitem{pdg2014} K. A. Olive etal. [Particle Data Group], \CHC38, 090001 (2014).


\bibitem{rb-exp} K. Nakamura et al. [Particle Data Group], \JPG37, 075021
(2010).

\bibitem{hrr1loop} A. Djouadi, \PR459, 1 (2008).

\bibitem{htm} W. Konetschny, W. Kummer, \PLB70, 433 (1977);
J. Schechter, J. W. F. Valle, \PRD 22, 2227 (1980); T. P. Cheng, L.
F. Li, \PRD22, 2860 (1980).

\bibitem{ihtm} T. Araki, C.Q. Geng, K. I. Nagao, \PRD83, 075014 (2011).

\bibitem{sepcomplex} T. Hambye, F.-S. Ling, L. Lopez Honorez, J. Rocher, \JHEP0907, 090 (2009) [Erratum-ibid. 1005, 066 (2010)];
F. S. Ling, arXiv:0905.4823; K. Earl, K. Hartling, H. E. Logan, T. Pilkington, \PRD88, 015002 (2013);
 C. Garcia-Cely, A. Ibarra, A. S. Lamperstorfer, M. H. G. Tytgat, arXiv:1507.05536; L. Wang, X.-F. Han, \PRD87, 015015 (2013);
\PRD86, 095007 (2012); Y. Cai, W. Chao, S. Yang, \JHEP1212, 043 (2012); Y. Hamada, K. Kawana, K. Tsumura, \PLB747, 238 (2015).

\bibitem{sepreal} C. Cai, Z.-M. Huang, Z. Kang, Z.-H. Yua, H.-H. Zhang, \PRD92, 115004 (2015).

\bibitem{atlas-1403.5294} ATLAS collaboration, \JHEP1405, 071 (2014).

\bibitem{cms-1405.7570} CMS collaboration, \EPJC74, 3036 (2014).

\bibitem{atlas-1402.7029} ATLAS collaboration, \JHEP1404, 169 (2014).

\bibitem{1602.00590} M. van Beekveld, W. Beenakker, S. Caron, R. R. de Austri, arXiv:1602.00590.

\bibitem{planck} P. A. R. Ade et al. [Planck Collaboration], \ASAS571, A16 (2014).


\bibitem{htmtheo} A. Arhrib, R. Benbrik, M. Chabab, G. Moultaka, M. C. Peyranere, L. Rahili, J. Ramadan, \PRD84, 095005 (2011).

\bibitem{1112.5453} A. Arhrib, R. Benbrik, M. Chabab, G. Moultaka, L. Rahili, \JHEP1204, 136 (2012).

\bibitem{yun} J. Bernon, J. F. Gunion, H. E. Haber, Y. Jiang, S. Kraml, \PRD92, 075004 (2015).


\end{thebibliography}
\end{document}